\documentclass{article}
\usepackage{amsmath}
\usepackage{arxiv}

\usepackage[utf8]{inputenc} 
\usepackage[T1]{fontenc}    
\usepackage{hyperref}       
\usepackage{url}            
\usepackage{booktabs}       
\usepackage{amsfonts}       
\usepackage{nicefrac}       
\usepackage{microtype}      
\usepackage{cleveref}       
\usepackage{lipsum}         
\usepackage{graphicx}
\usepackage{natbib}
\usepackage{doi}

\usepackage{listings}
\title{Application of GPU-accelerated FDTD method to electromagnetic wave propagation in plasma using MATLAB Parallel Processing Toolbox}

\author{ \href{https://orcid.org/0000-0002-8323-2290}{\includegraphics[scale=0.06]{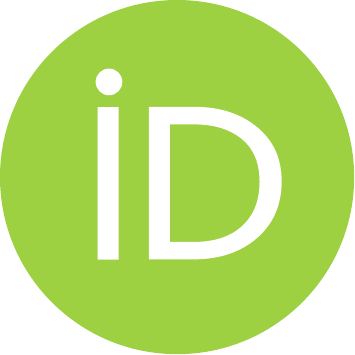}\hspace{1mm}Shayan~Dodge}\thanks{Use footnote for providing further
		information about author (webpage, alternative
		address)---\emph{not} for acknowledging funding agencies.} \\
	Laser and Plasma Reserch Insititute\\
	Shahid Beheshti University\\
	Iran\\
	\texttt{dodgeshayan@gmail.com} \\
	\And
	\href{https://orcid.org/0000-0000-0000-0000}{\hspace{0mm}Mojtaba Shafiee} \\
	Laser and Plasma Reserch Insititute\\
	Shahid Beheshti University\\
	Iran\\
	\texttt{m-shafiee@sbu.ac.ir} \\
	\AND
	\href{https://orcid.org/0000-0000-0000-0000}{\hspace{0mm}Babak Shokri} \\
    Laser and Plasma Reserch Insititute\\
    Shahid Beheshti University\\
    Iran\\
    \texttt{b-shokri@sbu.ac.ir} \\
}

\renewcommand{\shorttitle}{\textit{arXiv}}

\hypersetup{
pdftitle={the arxiv style},
pdfsubject={q-bio.NC, q-bio.QM},
pdfauthor={David S.~Hippocampus, Elias D.~Striatum},
pdfkeywords={First keyword, Second keyword, More},
}

\begin{document}
\maketitle

\begin{abstract}
Since numerical computing with MATLAB offers a wide variety of advantages, such as easier developing and debugging of computational codes rather than lower-level languages, the popularity of this tool is significantly increased in the past decade. However, MATLAB is slower than other languages. Moreover, utilizing MATLAB parallel computing toolbox on the Graphics Processing Unit (GPU) face some limitations. The lack of attention to these limitations reduces the program execution speed. Even sometimes, parallel GPU codes are slower than serial. 
In this paper, some techniques in using MATLAB parallel computing toolbox are studied to improve the performance of solving complex electromagnetic problems by the Finite Difference Time Domain (FDTD) method. Implementing these techniques allows the GPU-Accelerated Parallel FDTD code to execute 20x faster than (basic) serial FDTD code. 
Eventually, GPU-Accelerated Parallel FDTD code is utilized to optimize the computational modeling of electromagnetic waves propagating in plasma. In this simulation, kinetic theory equations for plasma are used (excluding inelastic collisions), and temporal evolution is studied by the FDTD method (coupled FDTD with kinetic theory).
\end{abstract}

\keywords{FDTD \and GPU \and Plasma \and Kinetic Theory \and Waves}

\section{Introduction}
The Finite Difference Time Domain (FDTD) method among the various numerical methods is the most widely used to solve electromagnetic problems, as it has a simple structure and high accuracy. In other to solving Maxwell’s equations to obtain the distribution of electric and magnetic fields in space and time domains, the FDTD method is the best approach. Maxwell’s equations are discretized using the Yee algorithm in this method, and the distribution of electric and magnetic fields in time and space is obtained.\citep{elsherbeni2015finite} However, when the mesh number or the number of time steps is very large, the execution speed decreases dramatically. Therefore, execution speed is a critical factor in this condition.

Traditional serial FDTD codes do not have enough performance. Hence the best way to increase the FDTD code execution speed is parallelization.\citep{elsherbeni2015finite} However, it is not an easy task because the explicit leap-frog scheme has been applied in the FDTD method.

Today, a Graphics Processing Unit (GPU) benefits from hundreds of computing cores with higher computing ability than a Central Processing Unit (CPU). Therefore, many attempts have been made to run programs on a GPU. \cite{yu2013valu,demir2010stacking,jiang2011cuda,xiong2018electromagnetic,diener2017fdtd,weiss2019using}.

In the past, parallelization of the FDTD method on a GPU using CUDA Kernels has been well developed.\citep{demir2010stacking,jiang2011cuda,xiong2018electromagnetic} Besides, many works have been done in optimizing GPU FDTD programming in MATLAB.\citep{diener2017fdtd,weiss2019using} However, there are still limitations that significantly impact the code execution speed.

Since MATLAB is widely used in most engineering fields, optimized MATLAB code is required. However, despite the advantages of programming over MATLAB, such as ease of use, MATLAB is slower than lower-level languages. So increasing the execution speed in the MATLAB environment is a very important factor.
FDTD method has various applications in solving electromagnetic problems, such as electromagnetic wave propagation in plasma. Since solving a system of equations is essential to simulate electromagnetic wave propagation in plasma, an optimized FDTD code is needed to execute this large number of computations.

There are several ways for plasma modeling using the FDTD method:

\begin{itemize}
	\item
The Drude model is commonly used to approximate the frequency behavior of plasma. In this case, plasma is modeled as a dispersive medium \citep{guo2018scattering}.
	\item
The fluid model describes plasma using macroscopic quantities. This model solves a set of Maxwell equations and velocity moments of the Boltzmann equation simultaneously\citep{arcese2017plasma}.
	\item
Finally, the last and most complete way to model plasma is the kinetic model based on solving the Boltzmann equation. In this approach, the temporal evolution of the electron distribution function is computed \citep{cerri2008fdtd,russo2010self,liu2018boltzmann}.
\end{itemize}

This work is organized into six sections. In section 2, the formulation of the FDTD method is studied. In section 3, the modified GPU FDTD code is introduced. In section 4, the validation of the modified FDTD code in the GPU is studied. In section 5, the three-dimensional propagation of electromagnetic waves in plasma is simulated using the FDTD method and Boltzmann equation. Finally, a summary and conclusion are presented in Section 6.

\section{FORMULATION OF FDTD METHOD}

The formulation of the FDTD method begins with Maxwell's equations
\begin{align}\label{equ1}
	\mu_{0} \frac{{\partial \vec H}}{{\partial t}}&= -\nabla  \times \vec E-\vec M ,\\
	\varepsilon_{0} \frac{{\partial \vec E}}{{\partial t}} &= \nabla  \times \vec H-\vec J,\notag
\end{align}
where $\vec M$ and $\vec J$ are the magnetic and electric current densities, respectively. Discretized Maxwell’s equations, using the Yee algorithm, are  as follows:
\begin{align}
	\left. {{E_x}} \right|_{i,j,k}^{n + 1} =& {\left. {C_{{E_x}}^{{E_x}}} \right|_{i,j,k}} \times \left. {{E_x}} \right|_{i,j,k}^n+ {\left. {C_{{H_z}}^{{E_x}}} \right|_{i,j,k}} \times \left[ {\left. {{H_z}} \right|_{i,j,k}^{n + {1 \mathord{\left/{\vphantom {1 2}} \right.
					\kern-\nulldelimiterspace} 2}} - \left. {{H_z}} \right|_{i,j - 1,k}^{n + {1 \mathord{\left/
					{\vphantom {1 2}} \right.\kern-\nulldelimiterspace} 2}}} \right]\\\notag
	+& {\left. {C_{{H_y}}^{{E_x}}} \right|_{i,j,k}} \times \left[ {\left. {{H_y}} \right|_{i,j,k}^{n + {1 \mathord{\left/{\vphantom {1 2}} \right.
					\kern-\nulldelimiterspace} 2}} - \left. {{H_y}} \right|_{i,j,k - 1}^{n + {1 \mathord{\left/
					{\vphantom {1 2}} \right.\kern-\nulldelimiterspace} 2}}} \right] + {\left. {C_{{J_x}}^{{E_x}}} \right|_{i,j,k}} \times \left. {{J_x}} \right|_{i,j,k}^{n + {1 \mathord{\left/{\vphantom {1 2}} \right.\kern-\nulldelimiterspace} 2}},\notag
\end{align}
where
\begin{align}
	{\left. {C_{{E_x}}^{{E_x}}} \right|_{i,j,k}} =& \frac{{2{\varepsilon _0} - \delta t{{\left. {\sigma _x^e} \right|}_{i,j,k}}}}{{2{\varepsilon _0} + \delta t{{\left. {\sigma _x^e} \right|}_{i,j,k}}}},\\
	{\left. {C_{{H_z}}^{{E_x}}} \right|_{i,j,k}} =& \frac{{2\delta t}}{{2{\varepsilon _0}\delta y + \delta t{{\left. {\sigma _x^e} \right|}_{i,j,k}}\delta y}},\\
	{\left. {C_{{H_y}}^{{E_x}}} \right|_{i,j,k}} =& \frac{{2\delta t}}{{2{\varepsilon _0}\delta z + \delta t{{\left. {\sigma _x^e} \right|}_{i,j,k}}\delta z}},\\
	{\left. {C_{{J_x}}^{{E_x}}} \right|_{i,j,k}} =& \frac{{2\delta t}}{{2{\varepsilon _0} + \delta t{{\left. {\sigma _x^e} \right|}_{i,j,k}}}},
\end{align}
In the same way, the other components of the electric and magnetic fields are discretized.

Given that computer storage is limited, FDTD problem space should be truncated by special boundary conditions that simulate electromagnetic waves propagating continuously beyond the calculation. Convolutional Perfectly Matched Layer (CPML) is implemented among various types of absorbing boundary conditions. The equation for the CPML region is
\begin{align}\label{equ4}
	\left. {{E_x}} \right|_{i,j,k}^{n + 1} =& {\left. {C_{{E_x}}^{{E_x}}} \right|_{i,j,k}} \times \left. {{E_x}} \right|_{i,j,k}^n\\\notag
	+&\left( {{{{{\left. {C_{{E_x}}^{{H_z}}} \right|}_{i,j,k}}} \mathord{\left/
				{\vphantom {{{{\left. {C_{{E_x}}^{{H_z}}} \right|}_{i,j,k}}} {{{\left. {\kappa _z^e} \right|}_{i,j,k}}}}} \right.
				\kern-\nulldelimiterspace} {{{\left. {\kappa _z^e} \right|}_{i,j,k}}}}} \right) \times \left[ {\left. {{H_z}} \right|_{i,j,k}^{n + {1 \mathord{\left/
					{\vphantom {1 2}} \right.
					\kern-\nulldelimiterspace} 2}} - \left. {{H_z}} \right|_{i,j - 1,k}^{n + {1 \mathord{\left/
					{\vphantom {1 2}} \right.
					\kern-\nulldelimiterspace} 2}}} \right]\\\notag
	+&\left( {{{{{\left. {C_{{E_x}}^{{H_y}}} \right|}_{i,j,k}}} \mathord{\left/
				{\vphantom {{{{\left. {C_{{E_x}}^{{H_y}}} \right|}_{i,j,k}}} {{{\left. {\kappa _y^e} \right|}_{i,j,k}}}}} \right.
				\kern-\nulldelimiterspace} {{{\left. {\kappa _y^e} \right|}_{i,j,k}}}}} \right) \times \left[ {\left. {{H_y}} \right|_{i,j,k}^{n + {1 \mathord{\left/
					{\vphantom {1 2}} \right.\kern-\nulldelimiterspace} 2}} - \left. {{H_y}} \right|_{i,j,k - 1}^{n + {1 \mathord{\left/{\vphantom {1 2}} \right.\kern-\nulldelimiterspace} 2}}} \right]\\\notag
	+&{\left. {C\psi _y^{{E_x}}} \right|_{i,j,k}} \times {\left. {\psi _y^{{E_x}}} \right|_{i,j,k}} + {\left. {C\psi _z^{{E_x}}} \right|_{i,j,k}} \times {\left. {\psi _z^{{E_x}}} \right|_{i,j,k}},
\end{align}
where
\begin{align}
	&\left. {\psi _y^{Ex}} \right|_{i,j,k}^{n + {1 \mathord{\left/{\vphantom {1 2}} \right.
				\kern-\nulldelimiterspace} 2}} = b_y^e\left. {\psi _y^{ex}} \right|_{i,j,k}^{n + {1 \mathord{\left/{\vphantom {1 2}} \right.
				\kern-\nulldelimiterspace} 2}} + a_y^e\left[ {\left. {{H_z}} \right|_{i,j,k}^{n + {1 \mathord{\left/{\vphantom {1 2}} \right.
					\kern-\nulldelimiterspace} 2}} - \left. {{H_z}} \right|_{i,j - 1,k}^{n + {1 \mathord{\left/
					{\vphantom {1 2}} \right.\kern-\nulldelimiterspace} 2}}} \right],\\
	&b_y^e = {e^{ - (\frac{{{\sigma _{\max }}{{({\rho  \mathord{\left/
									{\vphantom {\rho  \delta }} \right.\kern-\nulldelimiterspace} \delta })}^{{n_{cpml}}}}}}{{\kappa _y^e}} + \alpha _y^e)\frac{{\delta t}}{{\varepsilon o}}}},\\
	&a_y^e = \frac{{{\sigma _{\max }}{{({\rho  \mathord{\left/
							{\vphantom {\rho  \delta }} \right.\kern-\nulldelimiterspace} \delta })}^{{n_{cpml}}}}}}{{\delta y({\sigma _{\max }}{{({\rho  \mathord{\left/
							{\vphantom {\rho  \delta }} \right.\kern-\nulldelimiterspace} \delta })}^{{n_{cpml}}}}\kappa _y^e + \alpha _y^{pe}\kappa {{_y^e}^2})}}\left[ {b_y^e - 1} \right],\\
	&{\sigma _{\max }} = {\sigma _{factor}} \times \frac{{{n_{cpml}} + 1}}{{150\pi \delta y}},\\
	&\alpha _y^e(\rho ) = {\alpha _{\min }} + ({\alpha _{\max }} - {\alpha _{\min }})(1 - \frac{\rho }{\delta }).
\end{align}
Here, $n_{cpml}$ is the order of CPML, 	$\rho$ is boundary distance to the desired point, $\delta$ is the thickness of the CPML layer, $\sigma_{factor}$ is about $0.7$ to $1.5$, $\kappa_{\max}$ is about $5$ to $11$, $\alpha_{\max}$ is about $0$ to $0.05$, $n_{cpml}$ is about $2$ to $4$ and the thickness of boundary is $8$ cells.

To measure the execution speed of the FDTD method, the number of processed cells per second is calculated \cite{demir2010stacking}:
\begin{equation}\label{equ5}
	NMCPS = \frac{{({n_s}_{teps} \times {n_x} \times {n_y} \times {n_z})}}{{{t_s}}} \times {10^{ - 6}},
\end{equation}
where $NMCPS$ is the number of the cells processed per
second, ${n_s}_{teps}$ is the total number of time steps, ${t_s}$ is the total simulation time, and ${n_x}, {n_y}$ ans ${n_z}$ are the number of cells in problem space in $x,y$ and $z$ directions, respectively.

The system specifications used in the simulation are graphics card $GeForce{\rm{ }}GTX{\rm{ }}1050$ and CPU $IntelCorei7 - 7700HQ@2.80GHz$.
\begin{table}[!h]
	\caption{\label{Notations} The notation of the FDTD parameters used in the code.}
	\centering
	\footnotesize
	\begin{tabular}{@{}p{5cm}l}
		\hline\hline		
		Parameters~~&Notations\\\hline	
		$+x,+y,+z$~~&$xp,yp,zp$\\
		$-x,-y,-z$~~&$xn,yn,zn$\\
		${C_{{E_x}}^{{E_x}}}, {C_{{E_x}}^{{H_x}}}, {C_{{E_x}}^{{H_y}}}$~~&$Cexex, Cexhx, Cexhy$\\
		${\psi_{{y}}^{{E_x}}}, {\psi_{{y}}^{{E_z}}}$~~&$Psi\_ex\_yn, Psi\_ez\_yn$\\	
		${C\psi_{{y}}^{{E_x}}}, {C\psi_{{y}}^{{E_z}}}$~~&$CPsi\_ex\_yn, CPsi\_ez\_yn$\\
		${a_{{y}}^{{e}}}, {b_{{y}}^{{e}}}$~~&$cpml\_a\_yn, cpml\_b\_yn$\\\hline\hline
	\end{tabular}\\
\end{table}

\section{MODIFICATION OF GPU FDTD CODE USING MATLAB}
Nowadays, the GPU has played a significant role in accelerating numerical calculations. Using the MATLAB parallel processing toolbox for parallelization code has some limitations. The lack of attention to these limitations reduces the execution speed in some cases, even slower than the serial mode on the CPU unit.

The basic FDTD source code is generated from \citep{elsherbeni2015finite}, and in this paper mentioned source code is optimized and parallelized using MATLAB parallel processing toolbox. 

In the first step, using the MATLAB parallel processing toolbox, all the matrices in the code are defined on the GPU using the \textbf{gpuArray} command.

\subsection{Avoiding Array Indexing}
The significant factor that decreases the execution speed is the array indexing, which should be avoided as much as possible. Since array indexing causes wasting time in each iteration, it should be avoided, especially when the number of iterations is large. Therefore, finding a solution to this problem is required for optimizing code. For this purpose, some corrections should be made, such as unifying the dimension of the field matrices and their coefficient matrices and defining some new auxiliary matrices before the main loop.

\subsubsection{Defining fields and their coefficients}
Field matrices should be defined in the same size
\lstset{language=Matlab}  
\begin{lstlisting}
	Ex=zeros(nx+1,ny+1,nz+1,'gpuArray');
	Ey=zeros(nx+1,ny+1,nz+1,'gpuArray');
	Ez=zeros(nx+1,ny+1,nz+1,'gpuArray');
\end{lstlisting}

After defining the coefficients of the fields, their dimensions are unified by joining a zero matrix to the coefficient matrices as follows:
\lstset{language=Matlab}  
\begin{lstlisting}
	Cexex=gpuArray(cat(1,Cexex,zeros(1,ny+1,nz+1)));
	Cexhz=gpuArray(cat(1,Cexhz,zeros(1,ny+1,nz+1)));
	Cexhy=gpuArray(cat(1,Cexhy,zeros(1,ny+1,nz+1)));
\end{lstlisting}

\subsubsection{Definition of auxiliary matrices}
Since all field matrices are defined in the same size, there are additional arrays that should always be zero. Therefore, auxiliary matrices are defined so that by multiplying the auxiliary matrix by the corresponding field matrix, the additional arrays stay zero, and the main arrays do not change. The size of fields must be $(nx+1 \times ny + 1 \times nz + 1)$ and arrays $({\rm{ }}nx + 1{\rm{ }},{\rm{ }}:{\rm{ }},{\rm{ }}:{\rm{ }})$ are additional array so in the auxiliary matrix, all the arrays corresponding to the additional arrays are zero and the others are one.

\lstset{language=Matlab}  
\begin{lstlisting}
	Ex_A = ones(nx+1,ny+1,nz+1,'gpuArray');
	Ex_A(nx+1,:,:)=0;
\end{lstlisting}

\subsubsection{Definition of boundary condition coefficients}
By considering the location of each boundary, the 3D matrix of boundary condition coefficients is defined. The boundary condition coefficients related to the $E_x$ field at the $-y$ boundary are given as follows: 

Since coefficients $Psi\_ex\_yn$ and $CPsi\_ex\_yn$ are associated with the $Ex$ field, then additional arrays in these coefficients and $Ex$ are same.

\lstset{language=Matlab}  
\begin{lstlisting}
	Psi_ex_yn=cat(1,zeros(1,n_cpml_yn,nz+1),Psi_ex_yn);
	Psi_ex_yn=cat(2,Psi_ex_yn,zeros(nx+1,ny-n_cpml_yn,nz+1));
	Psi_ex_yn=gpuArray(cat(2,zeros(nx+1,1,nz+1),Psi_ex_yn));
	CPsi_ex_yn=cat(1,CPsi_ex_yn,zeros(1,n_cpml_yn,nz+1));
	CPsi_ex_yn=cat(2,CPsi_ex_yn,zeros(nx+1,ny-n_cpml_yn,nz+1));
	CPsi_ex_yn=gpuArray(cat(2,zeros(nx+1,1,nz+1),CPsi_ex_yn));
\end{lstlisting}

Since coefficients $cpml\_a\_e\_yn$ and $cpml\_b\_e\_yn$ are associated with the $Hz$ field, additional arrays in these coefficients and $Hz$ are same.

\lstset{language=Matlab}  
\begin{lstlisting}
	cpml_a_e_yn=cpml_a_e_yn.*ones(nx,n_cpml_yn,nz+1);
	cpml_a_e_yn=cat(2,cpml_a_e_yn,zeros(nx,ny-n_cpml_yn,nz+1));
	cpml_a_e_yn=cat(2,zeros(nx,1,nz+1),cpml_a_e_yn));
	cpml_a_e_yn=gpuArray(cat(1,cpml_a_ey_yn,zeros(1,nyp1,nz+1)));
	cpml_b_e_yn=cpml_b_e_yn.*ones(nx,n_cpml_yn,nz+1);
	cpml_b_e_yn=cat(2,cpml_b_e_yn,zeros(nx,ny-n_cpml_yn,nz+1));
	cpml_b_e_yn=cat(2,zeros(nx,1,nz+1),cpml_b_e_yn);
	cpml_b_e_yn=gpuArray(cat(1,cpml_b_e_yn,zeros(1,nyp1,nz+1)));
\end{lstlisting}

Since coefficients $cpml\_a\_A\_e\_yn$ and $cpml\_b\_A\_e\_yn$ are associated with the $Hx$ field, additional arrays in these coefficients and $Hx$ are same.

\lstset{language=Matlab}  
\begin{lstlisting}
	cpml_a_A_e_yn=cpml_a_e_yn.*ones(nx+1,n_cpml_yn,nz);
	cpml_a_A_e_yn=cat(2,zeros(nx+1,1,nz),cpml_a_x_e_yn);
	cpml_a_A_e_yn=gpuArray(cat(3,cpml_a_x_e_yn,zeros(nxp1,ny+1,1)));
	cpml_b_A_e_yn=cpml_b_e_yn.*ones(nx+1,n_cpml_yn,nz);
	cpml_b_A_e_yn=cat(2,cpml_b_x_e_yn,zeros(nx+1,ny-n_cpml_yn,nz));
	cpml_b_A_e_yn=cat(2,zeros(nx+1,1,nz),cpml_b_e_yn);
	cpml_b_A_e_yn=gpuArray(cat(3,cpml_b_x_e_yn,zeros(nx+1,ny+1,1)));
\end{lstlisting}

\subsubsection{Corrections the main loop}
The main loop is composed of two functions: the first function calculates the magnetic field, and the second function calculates the electric field using this magnetic field. In these two functions, two new series of matrices are defined in addition to the fields, their coefficients, and boundary conditions coefficients. The following is an example of these matrices:

\lstset{language=Matlab}  
\begin{lstlisting}
Ez_y_Zeros=(Ez(:,1:end-1,:),zeros(nx+1,1,nz+1,'gpuArray'));
Ez_y_Shift=(Ez(:,2:end,:),zeros(nx+1,1,nz+1,'gpuArray'));

Hz_y_Zeros=(zeros(nx+1,1,nz+1,'gpuArray'),Hz(:,2:end-1,:),...
zeros(nx+1,1,nz+1,'gpuArray'));
Hz_y_Shift=(zeros(nx+1,1,nz+1,'gpuArray'),Hz(:,1:end-2,:),...
zeros(nx+1,1,nz+1,'gpuArray'));
\end{lstlisting}

\begin{itemize}
	\item The $Update\_H$ function is defined as follows:\\
	Evolution of magnetic fields at the boundary
	
	\lstset{language=Matlab}  
	\begin{lstlisting}
Psi_hx_yn=cpml_b_my_yn.*Psi_hx_yn+cpml_a_my_yn.*(Ez_y_Shift-Ez_y_Zeros);
Hx = Hx+CPsi_hx_yn.*Psi_hx_yn;
Psi_hx_yp=cpml_b_my_yp.*Psi_hx_yp+cpml_a_my_yp.*(Ez_y_Shift-Ez_y_Zeros);
Hx=Hx+CPsi_hx_yp.*Psi_hx_yp;	
Psi_hx_zn=(cpml_b_mz_zn).*Psi_hx_zn+(cpml_a_mz_zn).*(Ey_z_Shift-Ey_z_Zeros);
Hx=Hx+CPsi_hx_zn.*Psi_hx_zn;	
Psi_hx_zp=cpml_b_mz_zp.*Psi_hx_zp+cpml_a_mz_zp.*(Ey_z_Shift-Ey_z_Zeros);
Hy=Hy+CPsi_hy_zp.*Psi_hy_zp;
	\end{lstlisting}
	
	Evolution of magnetic field
	\lstset{language=Matlab}  
	\begin{lstlisting}
Hx=((Chxhx.*Hx) +(Chxey.*(Ey_z_Shift-Ey_z_Zeros))+...
(Chxez.*(Ez_y_Shift-Ez_y_Zeros))).*Hx_A;
	\end{lstlisting}
	
	\item The $Update\_E$ function is defined as follows:\\
	Evolution of electric fields at the boundary
	
	\lstset{language=Matlab}  
	\begin{lstlisting}
Psi_ex_yn=cpml_b_ey_yn.*Psi_ex_yn+cpml_a_ey_yn.*(Hz_y_Zeros-Hz_y_Shift);
Ex=Ex+CPsi_ex_yn.*Psi_ex_yn;
Psi_exy_yp=cpml_b_ey_yp.*Psi_ex_yp+cpml_a_e_yp.*(Hz_y_Zeros-Hz_y_Shift);
Ex=Ex+CPsi_ex_yp.*Psi_ex_yp;
Psi_ex_zn=cpml_b_ez_zn.*Psi_ex_zn+cpml_a_ez_zn.*(Hy_z_Zeros-Hy_z_Shift);
Ex=Ex+CPsi_ex_zn.*Psi_ex_zn;
Psi_ex_zp=cpml_b_ez_zp.*Psi_ex_zp+cpml_a_ez_zp.*(Hy_z_Zeros-Hy_z_Shift);
Ex=Ex+CPsi_ex_zp.*Psi_ex_zp;
	\end{lstlisting}
	
	Evolution of electric fields
	
	\lstset{language=Matlab}  
	\begin{lstlisting}
Ex=(Cexex.*Ex+Cexhz.*(Hz_y_Zeros-Hz_y_Shift)+...
Cexhy.*(Hy_z_Zeros-Hy_z_Shift)).*Ex_A;
	\end{lstlisting}
	
\end{itemize}

\begin{figure}[ht]
	\centering
	\includegraphics[scale=0.30]{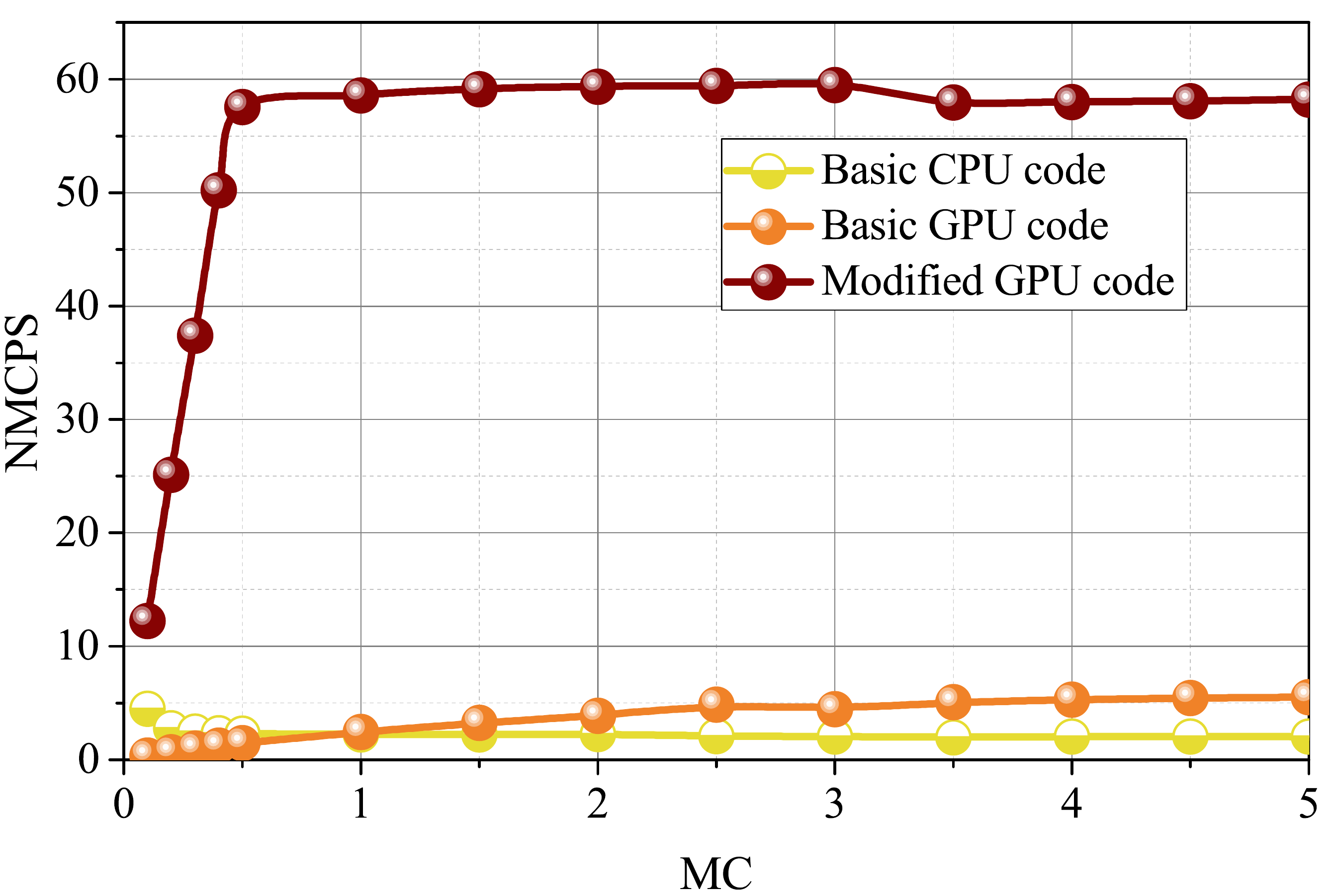}
	\caption{The execution speed of the basic and modified codes.}
	\label{CPU}
\end{figure}

As shown in Fig. \ref{CPU}, the GPU code execution speed for mesh number smaller than $1MC$ is slower than the CPU code. Only in larger than $2MC$, the GPU code execution speed is about three times faster than the basic CPU code. However, the modification can improve
execution speed up to 11 times.

\begin{table}[!h]
	\caption{\label{parameters} The auxiliary matrices are defined in the modified FDTD code.}
	\centering
	\footnotesize
	\begin{tabular}{@{}p{3.25cm}p{5.5cm}}
		\hline\hline		
		Parameters~~&Definition\\\hline
		$Ex\_A$&$Ex$ auxiliary matrix\\
		$cpml\_a\_A\_e\_yn$&$cpml\_a\_e\_yn$ auxiliary matrix\\
		$cpml\_b\_A\_e\_yn$&$cpml\_b\_e\_yn$ auxiliary matrix\\
		$Ez\_y\_Shift$&$Ez$ transformation matrix in $y$ direction\\
		$Ez\_y\_Zeros$&$Ez$ dimension matrix in $y$ direction\\\hline\hline
	\end{tabular}
\end{table}
\section{VALIDATION OF THE MODIFIED FDTD CODE IN THE GPU}
To validate the modified code, analytically and numerically solution of a Gaussian plane wave propagation in a vacuum is compared.

The Gaussian pulse is expressed as follows:
\begin{align}
	E\left( {x,t} \right) = \exp \left( { - {{\left( {\frac{{c\left( {t - {t_0}} \right) - x}}{{c\tau }}} \right)}^2}} \right),
\end{align}
where $t_0$ is the delay time and $\tau$ is the pulse width. In this case $\tau=0.5ns$ and $t_0=2.25ns$.

The normalized global error $L_2$ is calculated to evaluate the discretization error. If $u$ is the exact value of the function and ${\bar u}$ is the calculated result, the normalized global error between these two quantities is \citep{li2016verification}

\begin{align}
	{L_2} = \sqrt {\int {{{\left( {\bar u - u} \right)}^2}dx} }  = \sqrt {\frac{1}{N}\sum\limits_{i = 1}^N {{{\left( {{{\bar u}_i} - {u_i}} \right)}^2}} },
\end{align}
where N is the number of mesh. 

\begin{figure}[h]
	\centering
	\includegraphics[scale=0.25]{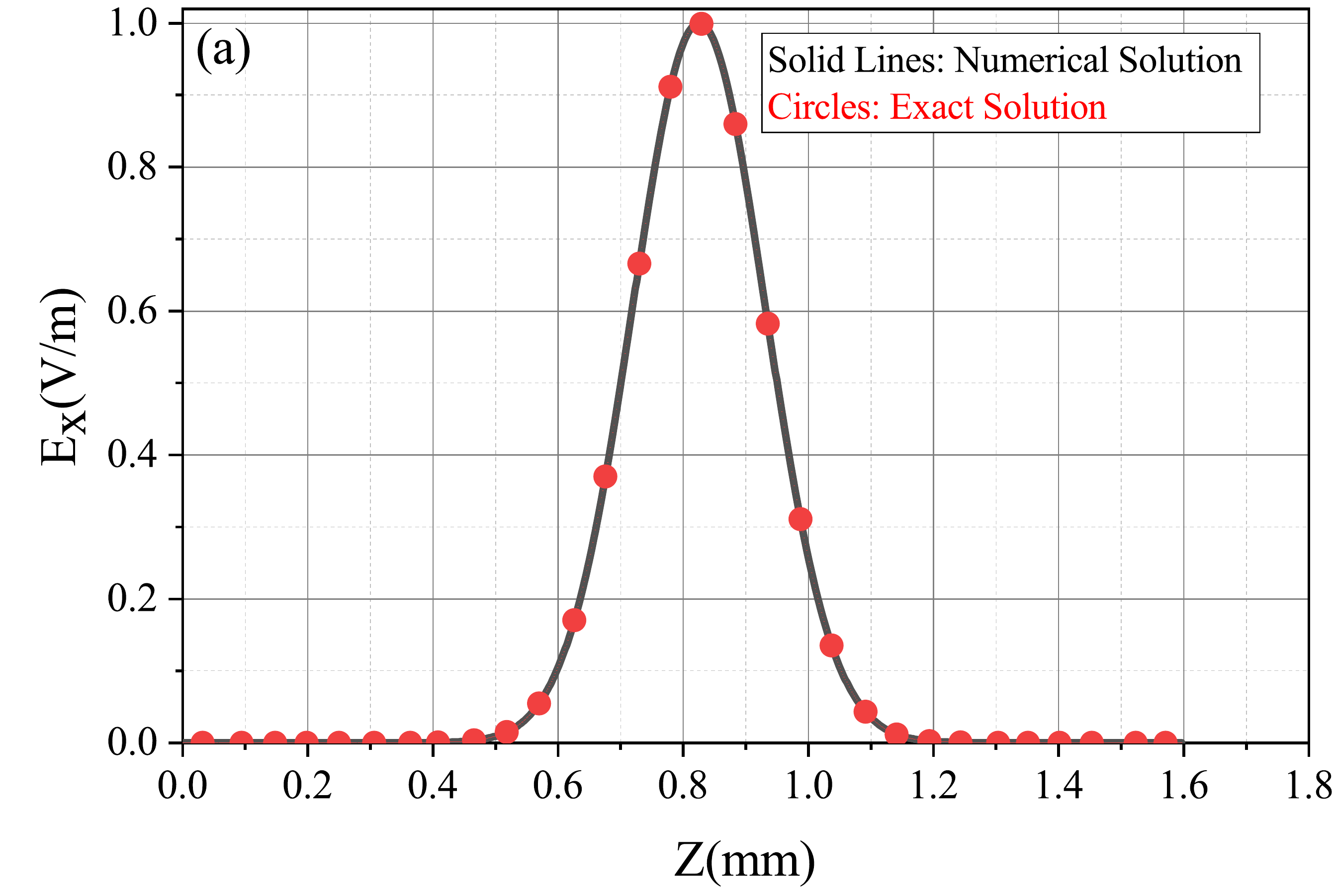}
	\includegraphics[scale=0.25]{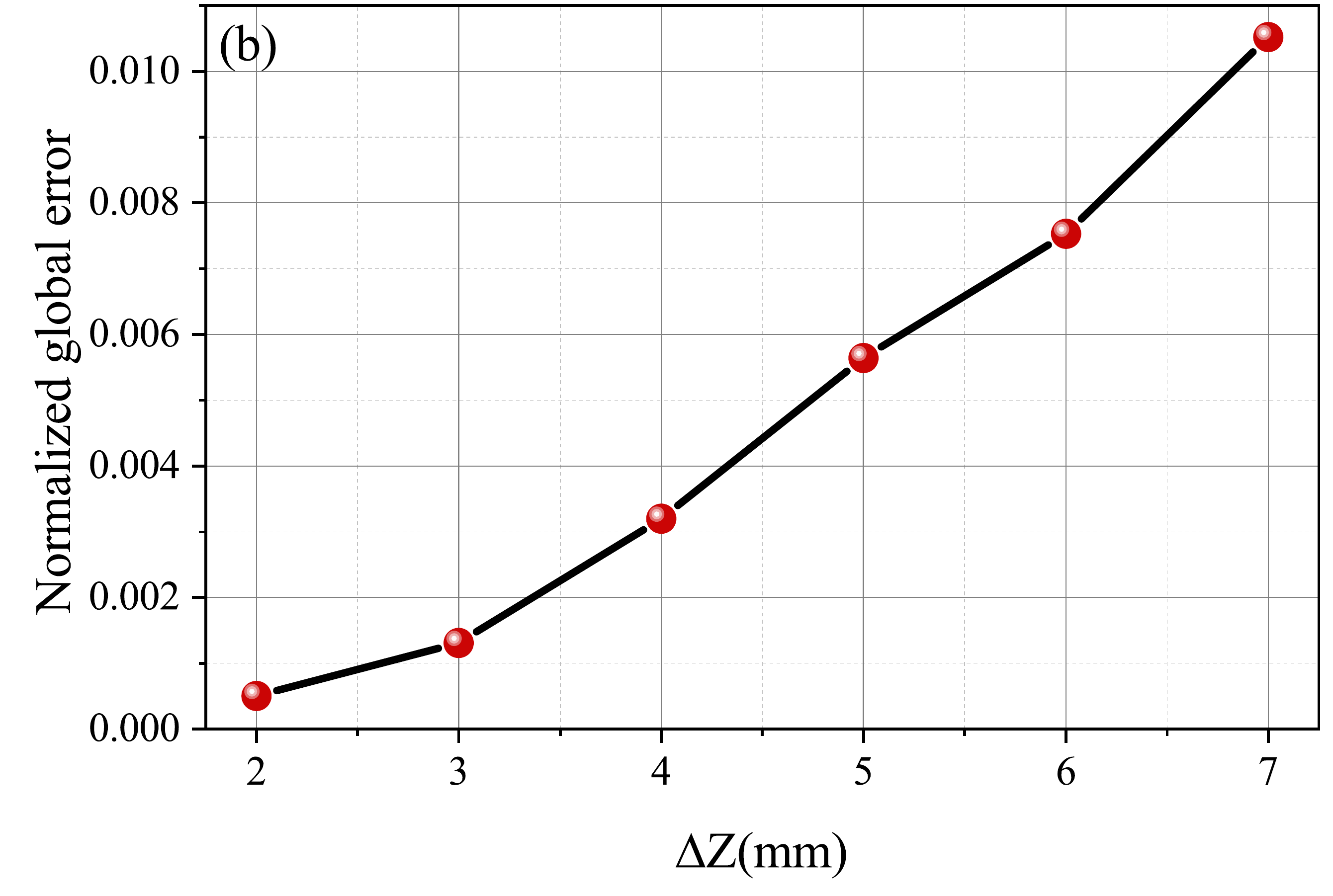}
	\caption{Validation for the Gaussian pulse (a) comparison between exact and numerical solutions calculated by the modified GPU code (b) the global error in various grid sizes.}
	\label{error}
\end{figure}
The exact and numerical results are significantly matched in Fig. \ref{error}(a). As shown in Fig. \ref{error}(b), the order of accuracy for the modified GPU code is 2.0, which is sufficient.
\section{PLASMA SIMULATION CODE IN THE GPU}
\subsection{Physics of the Problem}
The propagation of electromagnetic waves in plasma is studied by solving the Boltzmann and Maxwell equations system. The Boltzmann equation is
\begin{equation}\label{boltz}
	\frac{{\partial F}}{{\partial t}} + \mathord{\buildrel{\lower3pt\hbox{$\scriptscriptstyle\rightharpoonup$}} 
		\over v}  \cdot {\nabla _{\mathord{\buildrel{\lower3pt\hbox{$\scriptscriptstyle\rightharpoonup$}} 
				\over r} }}F + \frac{e}{m}\left( {\mathord{\buildrel{\lower3pt\hbox{$\scriptscriptstyle\rightharpoonup$}} 
			\over E}  + \mu \mathord{\buildrel{\lower3pt\hbox{$\scriptscriptstyle\rightharpoonup$}} 
			\over v}  \times \mathord{\buildrel{\lower3pt\hbox{$\scriptscriptstyle\rightharpoonup$}} 
			\over H} } \right) \cdot {\nabla _{\mathord{\buildrel{\lower3pt\hbox{$\scriptscriptstyle\rightharpoonup$}} 
				\over v} }}F = {\left[ {\frac{{\partial F}}{{\partial t}}} \right]_{coll}},
\end{equation}
where $F(\mathord{\buildrel{\lower3pt\hbox{$\scriptscriptstyle\rightharpoonup$}} 
	\over v} ,\mathord{\buildrel{\lower3pt\hbox{$\scriptscriptstyle\rightharpoonup$}} 
	\over r} ,t)$ is the electron distribution function.

The usual method to solve the Boltzmann equation is based on the expansion of the electron distribution function using spherical harmonics \cite{cerri2008fdtd}.
\begin{equation}\label{ex}
	{\mathord{\buildrel{\lower3pt\hbox{$\scriptscriptstyle\rightharpoonup$}} 
			\over F} }(v) = {F_{0}}(v) + \frac{\bar{v}}{v}\cdot\vec{{F_{1}}}(v),
\end{equation}
which ${F_0}(v)$ describes isotropic and
${\mathord{\buildrel{\lower3pt\hbox{$\scriptscriptstyle\rightharpoonup$}} \over F} _1}(v) = {F_{1x}}(v)\hat x + {F_{1y}}(v)\hat y + {F_{1z}}(v)\hat z$ describes non-isotropic motion of particles.

By substituting equation(\ref{ex}) into equation(\ref{boltz}) and after several manipulations based on the orthogonality of spherical harmonics, the following two equations for the isotropic and non-isotropic expressions are obtained:

\begin{align}
	&\frac{{\partial {F_0}}}{{\partial t}} + \frac{v}{3}{\nabla _{\vec r}} \cdot {\vec F_1} - \frac{e}{{3m}}\frac{{\mathord{\buildrel{\lower3pt\hbox{$\scriptscriptstyle\rightharpoonup$}} 
				\over E} }}{{{v^2}}} \cdot \frac{\partial }{{\partial v}}({v^2}{\kern 1pt} {\mathord{\buildrel{\lower3pt\hbox{$\scriptscriptstyle\rightharpoonup$}} 
			\over F} _1}) = {C_0}(F),\\
	&\frac{{\partial {{\vec F}_1}}}{{\partial t}} + v{\nabla _{\vec r}}{F_0} - \frac{e}{m}\mathord{\buildrel{\lower3pt\hbox{$\scriptscriptstyle\rightharpoonup$}} 
		\over E} \frac{{\partial {F_0}}}{{\partial v}} = {C_1}(F),
\end{align}	
where $C_0(F)$ and $C_1(F)$ are collision expressions. It is worth mentioning that in the present paper, only elastic collisions are considered.

\subsection{Numerical Model}
In this model, according to the experimental results, the isotropic electron distribution function is considered the Druyvesteyn function \citep{van2021arbitrary,li2018effects}.

The anisotropic electron distribution function is also considered zero at the initial moment, and its evolution is studied. Therefore, the three-dimensional model is described by the following system of equations:
\begin{align}\label{equ}
	&\mu_{0} \frac{{\partial \vec H}}{{\partial t}}=  -\nabla  \times \vec E-\vec M ,\\
	&\varepsilon_{0} \frac{{\partial \vec E}}{{\partial t}} =\quad \nabla  \times \vec H-\vec J,\\
	&\frac{{\partial {\kern 1pt} {{\mathord{\buildrel{\lower3pt\hbox{$\scriptscriptstyle\rightharpoonup$}} 
						\over F} }_1}}}{{\partial t}} - \frac{e}{m}\mathord{\buildrel{\lower3pt\hbox{$\scriptscriptstyle\rightharpoonup$}} 
		\over E} \frac{{\partial {F_0}}}{{\partial v}} =  - \upsilon (v){\kern 1pt} {\mathord{\buildrel{\lower3pt\hbox{$\scriptscriptstyle\rightharpoonup$}} 
			\over F} _1},
\end{align}
where the elastic collision frequency $\upsilon (v)$ and the current density $\vec J$ are as follows:
\begin{align}\label{constant collision frequency}
	&\upsilon (v)=Ns v,\\
	&\mathord{\buildrel{\lower3pt\hbox{$\scriptscriptstyle\rightharpoonup$}} 
		\over J}  =  - e\int_v {\mathord{\buildrel{\lower3pt\hbox{$\scriptscriptstyle\rightharpoonup$}} 
			\over v} F(\mathord{\buildrel{\lower3pt\hbox{$\scriptscriptstyle\rightharpoonup$}} 
			\over v} ,\mathord{\buildrel{\lower3pt\hbox{$\scriptscriptstyle\rightharpoonup$}} 
			\over r} ,t){d^3}v =  - e\frac{{4\pi }}{3}\int_0^\infty  {{v^3}{{\mathord{\buildrel{\lower3pt\hbox{$\scriptscriptstyle\rightharpoonup$}} 
						\over F} }_1}dv} } ,
\end{align}
where $s$ is the cross section for elastic collision between electrons and neutral atoms.

\begin{figure}[!h]
	\centering
	\includegraphics[scale=0.15]{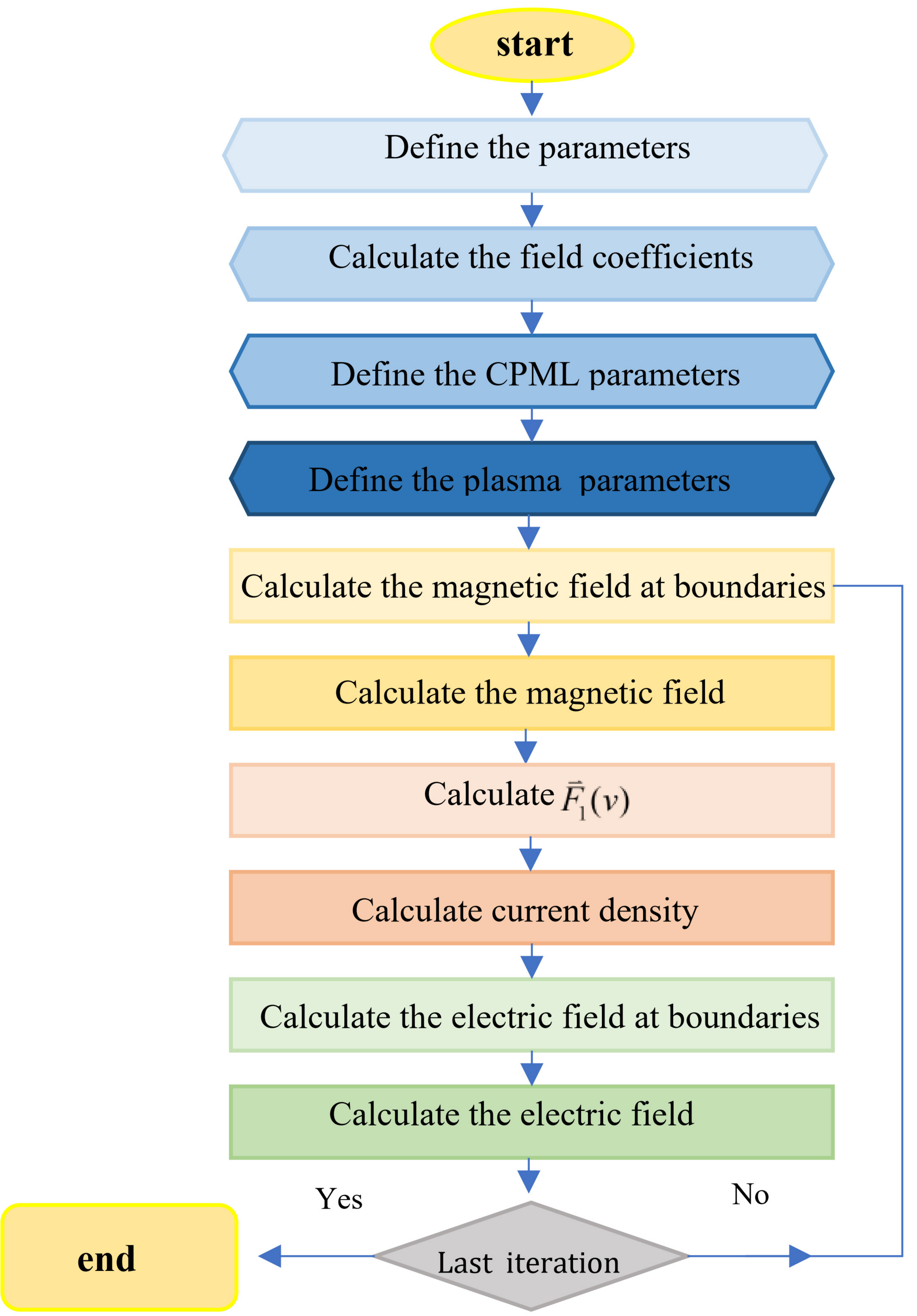}
	\caption{The flowchart used to simulate the propagation of electromagnetic	waves in plasma.}
	\label{alg2}
\end{figure}

The overall flowchart used to simulate the propagation of electromagnetic waves in plasma is shown in Fig. \ref{alg2}.

The anisotropic function is a four-dimensional parameter (three-dimensional space and one-dimensional velocity). Undoubtedly, calculating four-dimensional plasma parameters need sufficient execution speed, so optimizing the FDTD code's execution speed is vital. Therefore, the following modifications are applied. 

The anisotropic distribution function and the current density before the main loop should be defined as follows:
\begin{align}
	&f\_0 = zeros(nx + 1,ny + 1,nz + 1,v\_\dim ,'gpuArray');\\
	&f\_1x = zeros(nx + 1,ny + 1,nz + 1,v\_\dim ,'gpuArray');\\
	&Jx = zeros(nx + 1,ny + 1,nz + 1,'gpuArray');
\end{align}
where $v\_dim$ is the number of velocities considered. In this case $v\_dim = 31$. 

In the main loop, the anisotropic distribution function and the currents density are computed as follows
\begin{align}
	&f\_1x = (1./(2 + dt.*nu\_m)).*((2 - dt*nu\_m).*...\\\notag
	&f\_1x + (dt.*e./m\_e).*(def\_f\_0).*(2.*Ex));\\
	&Jx = ( - 4.*pi.*e./3).*gam.*sum(v^3.*f\_1x,4);
\end{align}
which $def\_f\_0$ is a derivative of the isotropic distribution function relative to velocity, and $gam$ is the velocity step.
\begin{table}[!h]
	\caption{\label{plasma_parameters} The utilized parameters for numerical study.}
	\centering
	\footnotesize
	\begin{tabular}{@{}p{10cm}p{1.5cm}}
		\hline\hline		
		Parameters
		&Amount\\\hline	
		Center frequency of the Gaussian-modulated pulse  ($f_c$)&$20~GHz$\\
		Bandwidth of the Gaussian-modulated pulse ($\Delta f$)&$20~GHz$\\
		Time constant of the Gaussian-modulated pulse ($\tau$)&$48~ps$\\
		Time shift of the Gaussian-modulated pulse ($t_0$)&$217~ps$\\
		Time Bandwidth of the Gaussian-modulated pulse ($\Delta t$)&$146~ps$\\
		Electron density ($n$)&$10^{19}~m^3$\\
		Neutral density  ($N$)&$10^{23}~m^3$\\
		Electron temperature ($T_e$)&$1~ev$\\\hline\hline	
	\end{tabular}\\
\end{table}
\subsection{Result}
The electromagnetic wave propagation in Argon plasma is considered to evaluate the performance of the modified plasma code. In this problem, the z-axis is divided into three regions. The two regions are air in  $z<0 mm$~and~$z>4 mm$, and the plasma region is located in $0<z<4 mm$. The incident plane wave is a Gaussian modulated, irradiated from air into plasma.

Plasma and electromagnetic waves properties are presented
in Table \ref{plasma_parameters}.

\begin{figure}[!h]
	\centering
	\includegraphics[scale=0.28]{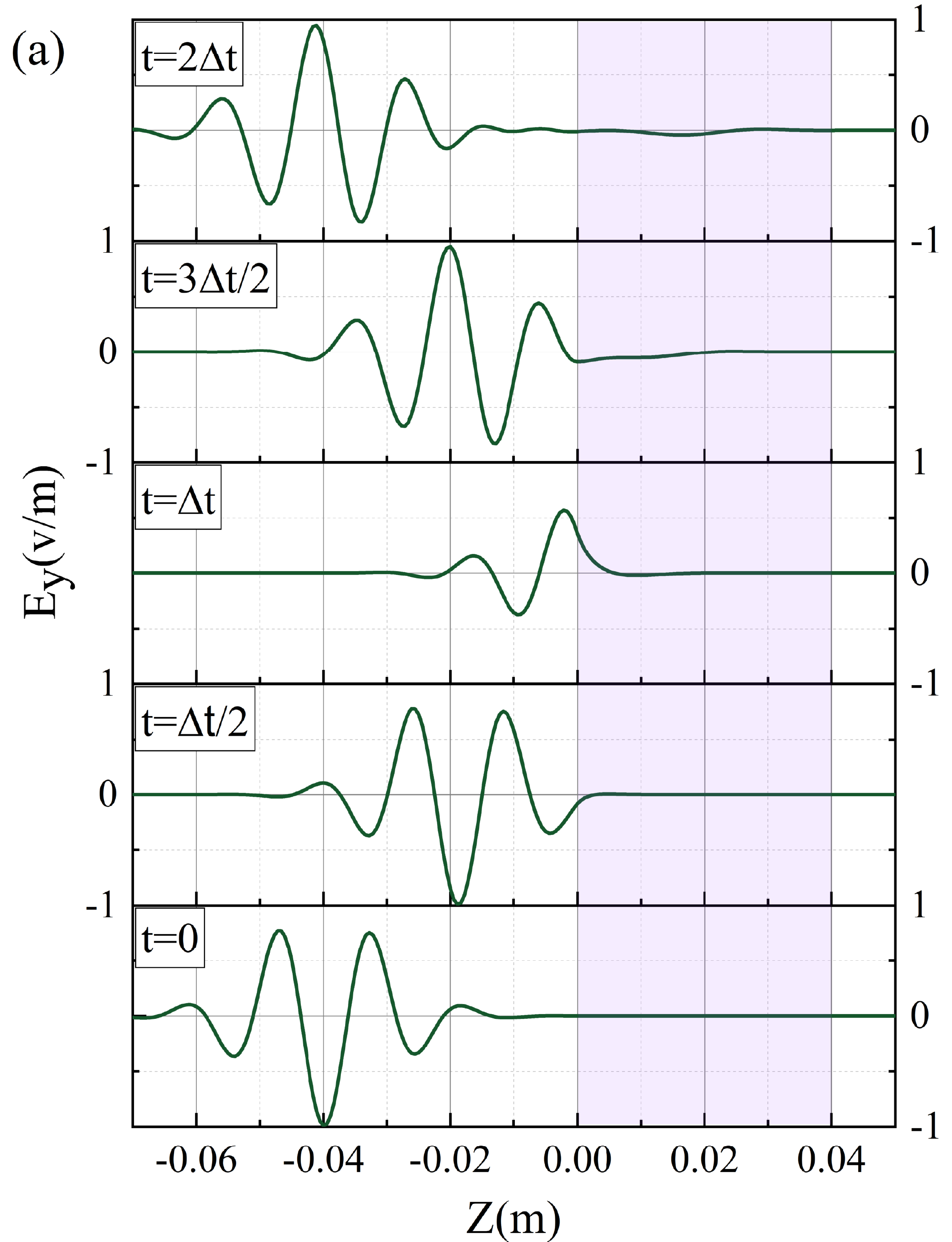}
	\includegraphics[scale=0.28]{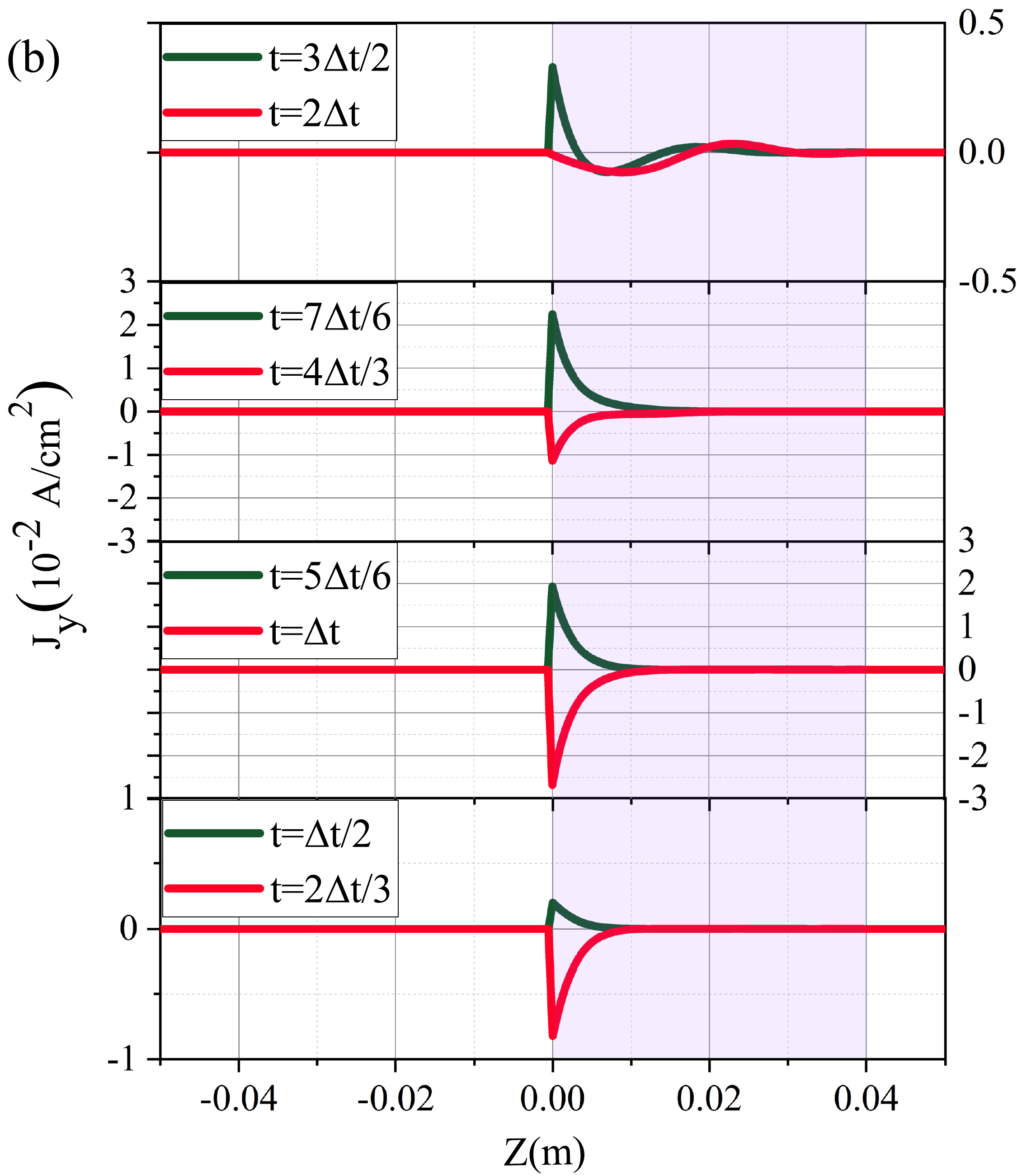}
	\caption{Temporal evolution of (a) the Gaussian-modulated plane wave radiation into plasma and (b) the plasma current density at the center of air-plasma interface along the z-axis.}
	\label{Ey5}
\end{figure}

Fig. \ref{Ey5}(a) shows two-dimensional views of the electric field evolution along the z-axis at different times. As shown in Fig. \ref{Ey5}(a), the Gaussian-modulated plane wave, which contains a frequency range, is irradiated to the plasma-air interface. Plasma behaves like a filter, so the waves with a frequency higher than the plasma frequency can be propagated into plasma, and the waves with a frequency lower than the plasma frequency are reflected from its interface. In this work, the electron-neutral cross-section described in reference \citep{alves2014lisbon} for cold Argon plasma is used.

According to the Figs. \ref{Ey5}(a) and (b) in the initial time steps ($t=0$ to $t=\Delta t/2$), there is no interaction between the electric field and plasma, so the current density and the electric field inside plasma are zero. Since the electric field is irradiated into the air-plasma interface from $t=\Delta t/2$, the plasma current density corresponding to the electric field on the interface continues to rise until it reaches a peak of $-30~mA/{cm}^2$ at the $t=\Delta t$ and then falls gradually until the electric field gets out of plasma ($t=3\Delta t/2$).

\begin{figure*}[!t]
	\centering
	\includegraphics[scale=0.32]{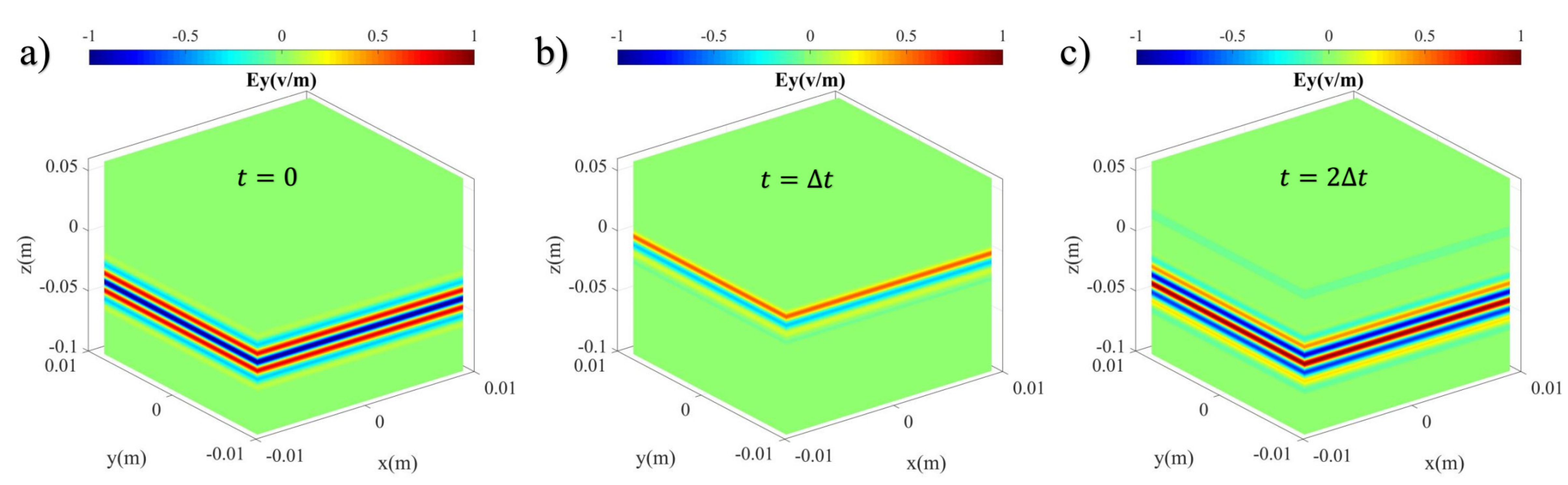}
	\caption{Propagation of the Gaussian-modulated plane wave into air and plasma.}
	\label{E3D}
\end{figure*}

In Fig. \ref{E3D}, the three-dimensional profile of the electric field is shown, (a) before the interface, (b) in the absorption region, (c) after the interaction (the high-frequency part is propagated, and the low-frequency part is reflected).
\begin{figure}[h]
	\centering
	\includegraphics[scale=0.30]{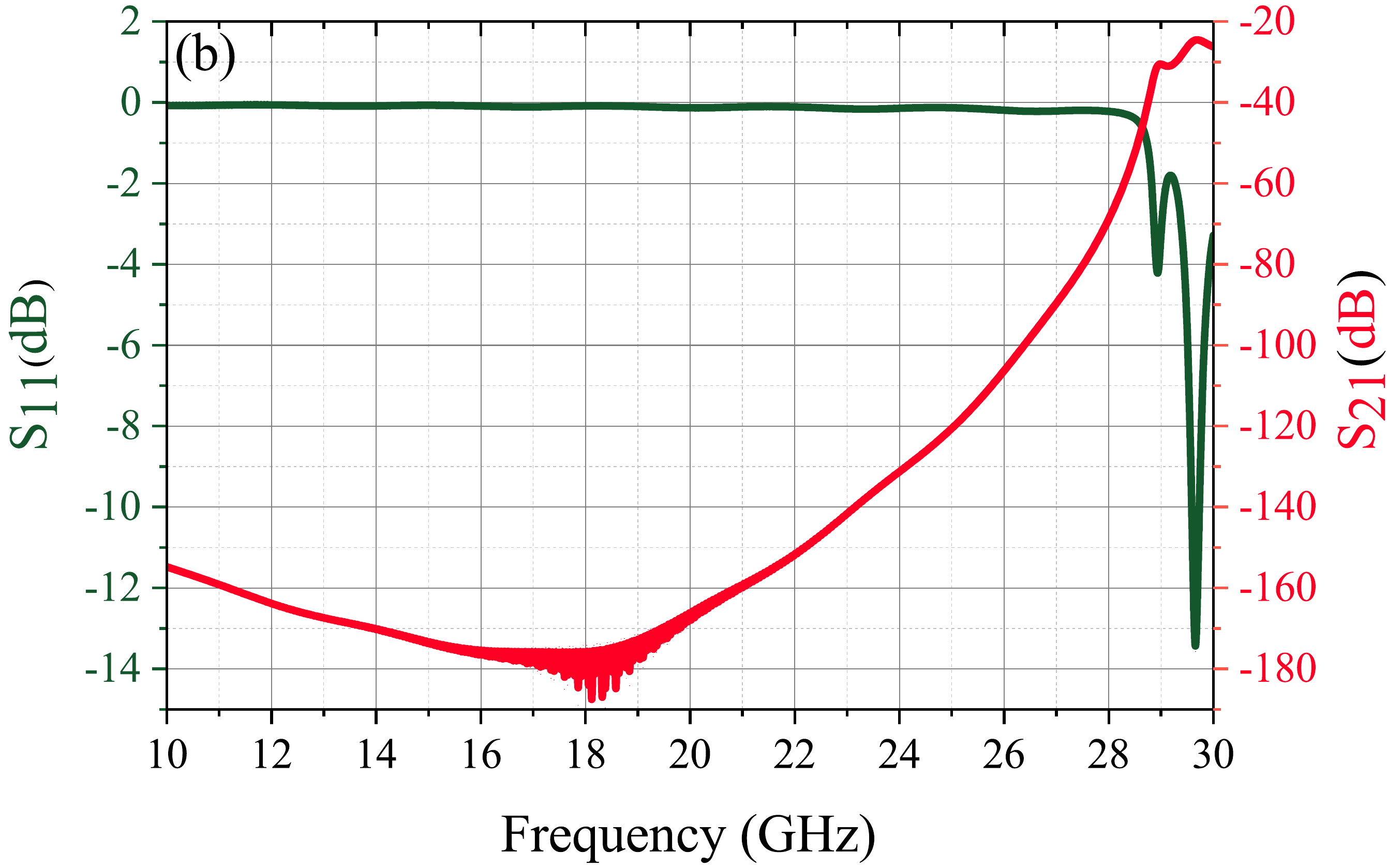}
	\caption{Scattering parameters for the plasma slab.}
	\label{f}
\end{figure}

Some scattering parameters, such as $S_{11}$ and $S_{21}$, were studied to investigate the reflection, absorption, and transmission of electromagnetic waves in plasma. These parameters show the reflection and transmission from plasma, respectively. Therefore, two ports are defined at the beginning ($z=0mm$) and end ($z=4mm$) of plasma to measure these two scattering parameters.

As shown in Fig. \ref{f}, plasma completely reflects the electromagnetic waves at the frequencies below the plasma frequency ($28.5GHz$). So in these frequencies, $s_{11}$ is zero. Moreover, plasma behaves like a lossy medium at the frequencies above the plasma, and a fraction of the wave is reflected, transmitted, and absorbed by the plasma. 

\begin{figure}[!h]
	\centering
	\includegraphics[scale=0.26]{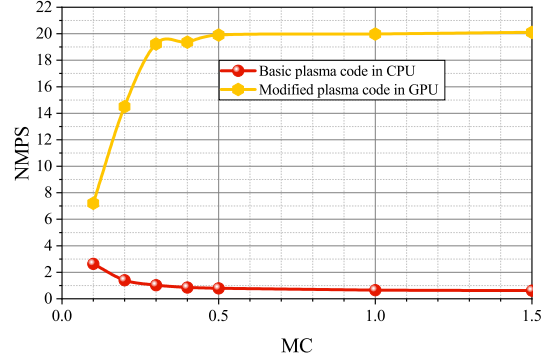}
	\caption{Comparison of the plasma basic code in CPU and modified plasma
		code in GPU.}
	\label{plasmas}
\end{figure}

The mentioned results described the physical behavior of plasma exactly. Furthermore, the execution speed of the modified GPU code is presented in Fig. \ref{plasmas}. This figure shows that the plasma simulation execution speed significantly increased(up to 20 times) using GPU and attention to the mentioned modifications.\\

The modified codes' significant advantages are summarized in Table \ref{conclusion}.

\begin{table}[h]
	\caption{\label{conclusion} Summary of significant advantages of mentioned modifications.}
	\centering
	\footnotesize
	\begin{tabular}{@{}p{4.5cm}p{5.5cm}}
		\hline\hline		
		Modification
		&Advantages\\\hline	
		Modified FDTD GPU code& $11x$ faster than basic GPU code\\
		Modified plasma code in GPU& $20x$ faster than basic plasma code in CPU\\\hline\hline	
	\end{tabular}\\
\end{table}

\section{Conclusion}
Several programming techniques were presented to reduce the execution time of the FDTD simulation in MATLAB software. Using MATLAB parallel computing toolbox and mentioned techniques, FDTD code was optimized on GPU. Validation of the modified code was examined, and the desired order of accuracy for the modified code was shown. Eventually, the interaction of electromagnetic waves with plasma (modeled by two-term Boltzamnn approximation) was simulated on the GPU by applying the proposed techniques, and its significant effect on the execution speed of the plasma code was proved.

\bibliographystyle{unsrtnat}
\bibliography{references}  
\end{document}